\documentclass[pra,twocolumn,nofootinbib,floatfix,10pt]{revtex4-2}
\usepackage{textcomp,mathcomp}
\usepackage{amsmath}
\usepackage{amssymb}
\usepackage{wasysym}
\usepackage{graphicx}
\usepackage{color,soul}
\usepackage{physics}
\usepackage{siunitx}
\usepackage{dsfont}
\usepackage{float}
\usepackage[english]{babel}
\usepackage{blindtext}
\usepackage[english,nomargin,inline,marginclue,draft]{fixme}
\pdfpageheight\paperheight
\pdfpagewidth\paperwidth

\usepackage[colorlinks,linkcolor=blue,anchorcolor=blue,citecolor=blue,urlcolor=blue]{hyperref}

\fxusetheme{colorsig}
\FXRegisterAuthor{cg}{acg}{CG}  
\FXRegisterAuthor{th}{ath}{\color{blue}TH}  
\FXRegisterAuthor{ib}{aib}{\color{red}IB} 
\FXRegisterAuthor{sh}{ash}{\color{cyan}SH} 
\FXRegisterAuthor{db}{adb}{\color{green}DB} 
\FXRegisterAuthor{ps}{aps}{PS}
\makeatletter
\renewcommand*\FXLayoutInline[3]{%
  {\@fxuseface{inline}\ignorespaces{\color{fx#1}[#3: #2]}}}
\makeatother

\long\def\symbolfootnote[#1]#2{\begingroup%
\def\thefootnote{\fnsymbol{footnote}}\footnotetext[#1]{#2}\endgroup}

\def\nobreakbefore{%
  \relax\ifvmode\else
    \ifhmode
      \ifdim\lastskip > 0pt\relax
        \unskip\nobreakspace
      \else 
        \nobreakspace
      \fi
    \fi
  \fi
}
\let\oldcite\cite
\renewcommand\cite{\nobreakbefore\oldcite}

\begin{document}
\title{Observation of  Discrete Time Quasicrystal in Rydberg Atomic Gases}

\author{Dong-Yang Zhu$^{1,2,\textcolor{blue}{\star}}$}
\author{Zheng-Yuan Zhang$^{1,2,\textcolor{blue}{\star}}$}
\author{Qi-Feng Wang$^{1,2,\textcolor{blue}{\star}}$}
\author{Yu Ma$^{1,2}$}
\author{Tian-Yu Han$^{1,2}$}
\author{Chao Yu$^{1,2}$}
\author{Qiao-Qiao Fang$^{1,2}$}
\author{Shi-Yao Shao$^{1,2}$}
\author{Qing Li$^{1,2}$}
\author{Ya-Jun Wang$^{1,2}$}
\author{Jun Zhang$^{1,2}$}
\author{Han-Chao Chen$^{1,2}$}
\author{Xin Liu$^{1,2}$}
\author{Jia-Dou Nan$^{1,2}$}
\author{Yi-Ming Yin$^{1,2}$}
\author{Li-Hua Zhang$^{1,2}$}
\author{Guang-Can Guo$^{1,2}$}
\author{Bang Liu$^{1,2,\textcolor{blue}{\ddagger}}$}
\author{Dong-Sheng Ding$^{1,2,\textcolor{blue}{\dagger}}$}
\author{Bao-Sen Shi$^{1,2}$}

\affiliation{$^1$Laboratory of Quantum Information, University of Science and Technology of China, Hefei, Anhui 230026, China.}
\affiliation{$^2$Anhui Province Key Laboratory of Quantum Network, University of Science and Technology of China, Hefei 230026, China.}

\date{\today}
\symbolfootnote[1]{D.Y.Z, Z.Y.Z, Q.F. W contribute equally to this work.}
\symbolfootnote[3]{lb2016wu@ustc.edu.cn}
\symbolfootnote[2]{dds@ustc.edu.cn}

\maketitle

\textbf{Discrete time quasicrystals (DTQC) constitute a class of non-equilibrium matter characterized by temporal order without strict periodicity, in contrast to conventional time crystals. Investigating these phenomena is essential for expanding our fundamental understanding of far-from-equilibrium quantum matter and spontaneous symmetry breaking beyond periodic regimes. Here, we experimentally observe a DTQC in a driven-dissipative ensemble of strongly interacting Rydberg atoms, displaying non-equilibrium dynamical response with a different finite Abelian group symmetry $\mathbb{Z}{_m} \times \mathbb{Z}{_n}$. By applying a quasi-periodic drive using a dual-frequency drive with incommensurate frequencies, we demonstrate that the system exhibits a robust subharmonic response at multiple incommensurate frequencies, signifying the emergence of a DTQC phase. We map the full phase diagram of the system, which includes the DTQC phase, and demonstrated its rigidity against perturbations in both RF field intensity and laser detuning. Moreover, we observe a cyclic group symmetry effect that constrains the construction of $\mathbb{Z}{_2} \times \mathbb{Z}{_3}$-symmetric DTQC. This work establishes a versatile platform for studying non-equilibrium phases of matter and provides insights into the dynamics of time-translation symmetry breaking in quantum many-body systems. }

\section*{INTRODUCTION}
{Spontaneous symmetry breaking is a fundamental concept in physics science and plays a crucial role in understanding phases of matter and phase transitions \cite{chaikin1995principles,sachdev1999quantum}. In recent years, increasing attention has been directed toward the spontaneous breaking of time-translation symmetry, giving rise to the concept of time crystals. Originally proposed by Frank Wilczek \cite{wilczek2012quantum}, time crystals have since been extensively investigated both theoretically and experimentally in different physical systems \cite{zhang2017observation,choi2017observation,watanabe2015absence,syrwid2017time,huang2018clean,gong2018discrete,yao2020classical,li2012space,else2016floquet,PhysRevA.91.033617,autti2018observation,smits2018observation,pizzi2021bistability,autti2021ac,trager2021real,kyprianidis2021observation,kucsko2018critical,bruno2013impossibility,taheri2022all}. In general, time crystals can be divided into discrete and continuous time crystals \cite{sacha2017time,else2020discrete,kongkhambut2022observation,zaletel2023colloquium,sacha2020time} that exhibit discrete and continuous time-translation symmetry breaking, respectively. While conventional time crystals exhibit strict periodicity and long-range order, time quasicrystals constitute a notable exception: they lack exact periodicity yet retain long-range order\cite{else2020long,PhysRevB.100.134302}, which is emerging from the spontaneous breaking of more complex time-translation symmetry \cite{zaletel2023colloquium,PhysRevX.15.011055}. The discovery of quasicrystals has significantly expanded our understanding of the structural nature of matter and has opened up a wide range of potential applications \cite{PhysRevLett.53.2477,PhysRevB.34.596,steinhardt1987physics}. Analogously, the concept of time quasicrystals has emerged as an important research direction \cite{PhysRevX.15.011055,autti2018observation,pizzi2019period,PhysRevResearch.6.023054,PhysRevLett.120.140401,PhysRevB.100.134302,giergiel2019discrete}. Investigating time quasicrystals is expected to provide deeper insights into the non-equilibrium dynamics of quantum many-body systems.

Owing to the strong interactions between Rydberg atoms, driven-dissipative Rydberg ensembles provide an excellent experimental platform for exploring non-equilibrium phenomena, including self-organization and non-equilibrium phase transitions\cite{lee2012collective,carr2013nonequilibrium,helmrich2020signatures,ding2019Phase,ding2022enhanced,wadenpfuhl2023emergence,ding2023ergodicity}. Furthermore, dissipative time crystals\cite{wu2023observation,liu2024bifurcation}, higher-order discrete time crystals (DTCs), and fractional-order DTCs \cite{liu2024higher} have been experimentally realized. Under the application of a periodic external drive, the system exhibits subharmonic responses, thereby breaking discrete time-translation symmetry. When subjected to a quasi-periodic drive\cite{he2023quasi,abanin2015exponentially,PhysRevB.97.020303,PhysRevLett.120.070602}, the system departs from equilibrium, and the interplay between Rydberg-atom interactions and dissipation leads to the emergence of a time quasi-crystal phase, offering a unique platform for studying non-equilibrium dynamics in a more complex finite Abelian group symmetry.

\begin{figure*}
\centering
\includegraphics[width=2\columnwidth]{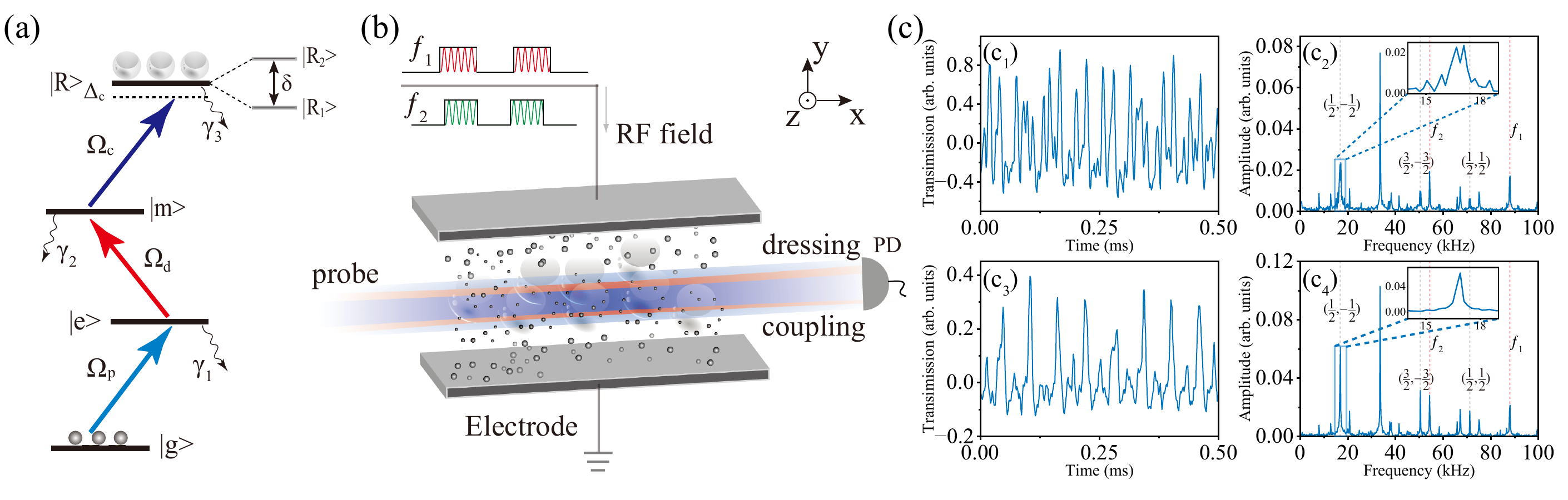}\\
\caption{\textbf{Experimental setup and time quasi-crystal order observations.} (a) Energy level diagram based on the three-photon Rydberg electromagnetically induced transparency (EIT) scheme. $\gamma _{1} $, $\gamma _{2} $, and $\gamma _{3} $ correspond to the decoherence rates of states $\left |e \right \rangle $, $\left |m \right \rangle $, and $\left |R \right \rangle $, respectively. When the atoms is driven by a radio-frequency field, the Rydberg level $\left |R \right \rangle $ generates a series of Floquet sidebands; the diagram shows two sideband levels $\left |R_{1} \right \rangle $ and $\left |R_{2} \right \rangle $, with an energy separation of $\delta $. (b) Simplified experimental setup. Two radio-frequency fields are applied to the atoms through a pair of electrodes with two channels (channel 1 and 2), with tunable frequencies. (c) The modulation frequency of the RF field of channel 1 is set to $f_{1} $. To achieve quasi-periodic behavior, the modulation frequency $f_{2} $ of channel 2 is configured with the maximally incommensurate ratio $r=f_{1} /f_{2} =(\sqrt{5}+1)/2$. (c1)-(c2) The time-domain and frequency-domain diagrams of the probe light transmission spectrum at an RF carrier frequency of 7.2 MHz, with a modulation frequency of 88 kHz in $f_1$-field and 54.387 kHz in $f_2$-field. Gray dashed lines labeled ($N_1/{m}$, $N_2/{n}$) denote the subharmonic response frequencies of the DTQC. Red dashed lines mark the quasiperiodic driving frequencies. Inset: Expanded subharmonic response at (1/2, -1/2).
}
\label{setup}
\end{figure*}

In this work, we experimentally observe discrete time quasicrystals (DTQCs) in a strongly interacting Rydberg atomic system. Utilizing a dual-frequency quasi-periodic drive, we observe a characteristic subharmonic response of the system, manifested as a linear combination of half the two drive frequencies, revealing the non-equilibrium many-body dynamics with distinct $\mathbb{Z}{_m} \times \mathbb{Z}{_n}$-symmetries and demonstrating the constrained effect for a cyclic group symmetry of $\mathbb{Z}{_2} \times \mathbb{Z}{_3}$. By tuning system parameters, the system exhibits a phase transition from a state with unbroken discrete-time symmetry to a time quasicrystal. We also show that the time quasicrystal exhibits robustness against perturbations in experimental parameters. These results not only advance our understanding of time crystals as a unique state of matter but also open new avenues for exploring and manipulating non-equilibrium dynamics in quantum many-body systems.}

\begin{figure*}
\centering
\includegraphics[width=2.08\columnwidth]{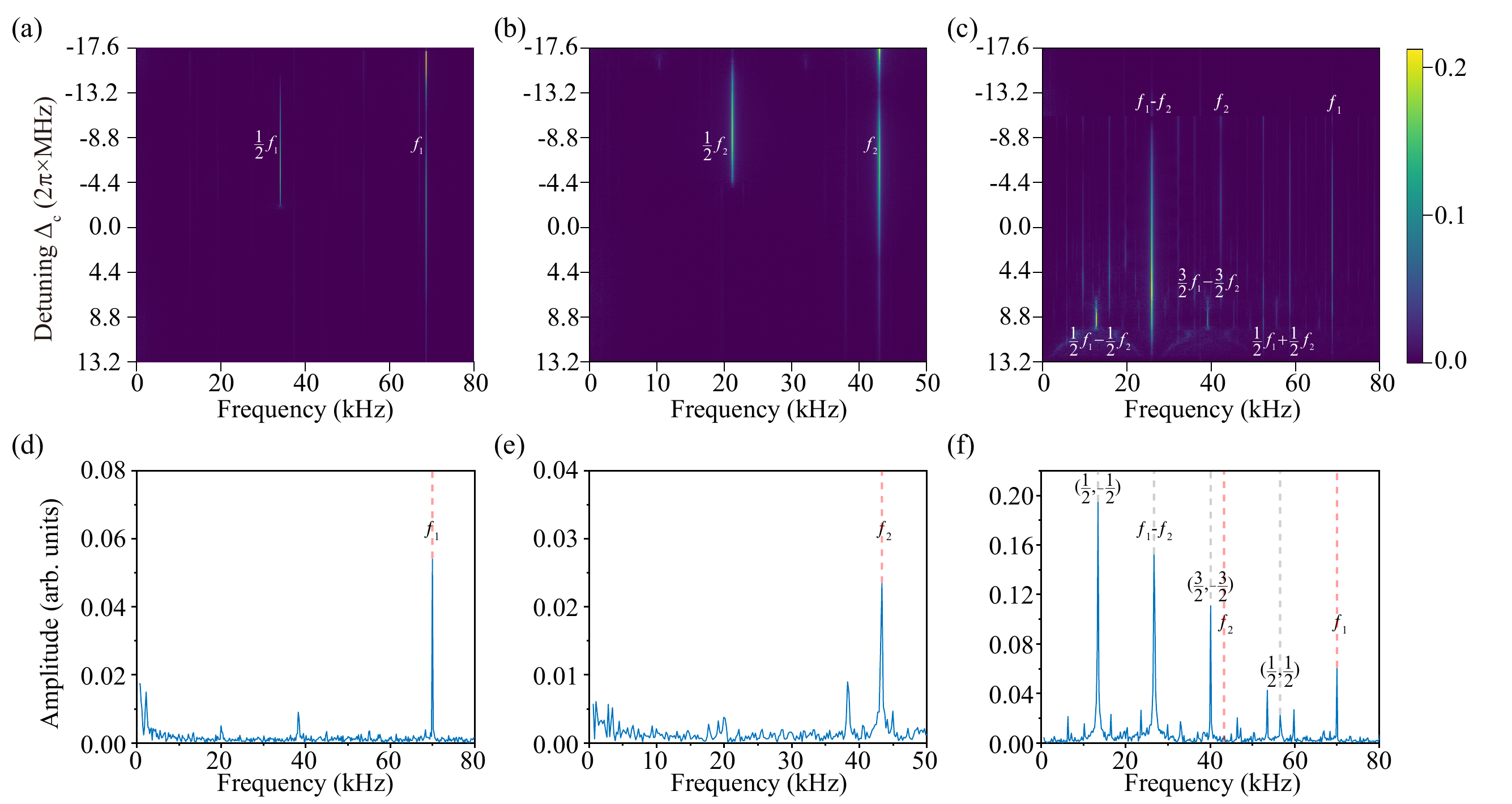}\\
\caption{\textbf{Measured phase diagrams under single- and dual-frequency driving.} Color maps show the transmission of the probe light during the scanning process. The coupling detuning $\Delta _{c}$ is scanned from $\Delta _{c} = - 2\pi \times 17.6$ MHz to $\Delta _{c} =  2\pi \times 13.2$ MHz. The frequency of both RF fields is set to 7.2 MHz. (a) shows the measured result with only $f_1$-field turned on, where the voltage is set to $U_{1} $ = 1.7 V and the modulation frequency is $f_1$ = 70 kHz, while $f_2$-field remains off. (b) shows the result with $f_1$-field turned off and $f_2$-field turned on, with a voltage $U_{2} $ = 1.5 V and modulation frequency $f_{2} $ = 43.262 kHz. (c) displays the phase diagram obtained when both two fields are turned on simultaneously. The color bar indicates the transmission intensity. (d)–(f) present the corresponding transmission spectra at a coupling detuning of $\Delta _{c} =  2\pi \times 8.8$ MHz for the measurements.}
\label{phase diagram}
\end{figure*}

\section*{RESULTS}
\subsection*{Physical model}
Our theoretical model is based on a three-level Rydberg atom system with one ground state $\left| g \right\rangle$ and two Rydberg states $\left|  R_1 \right\rangle$ and $\left|  R_2 \right\rangle$. The Hamiltonian of the system is based on a dual-frequency periodically driving double Rydberg state model \cite{wu2023observation,liu2024higher}: 
\begin{equation}
\begin{aligned}
    \hat{H}(t) & =\frac{1}{2}\sum_{i}\left(\Omega_{1}\sigma_{i}^{gR_1}+\Omega_{2}\sigma_{i}^{gR_2}+h.c.\right)\\ &-\sum_{i}\left[(\Delta_{f_1}(t)+\Delta_{f_2}(t) )(n_{i}^{R_1}+n_{i}^{R_2})+\delta n_{i}^{R_2}\right] \\ &+\sum_{i\neq j}V_{ij}\bigg[n_{i}^{R_1}n_{j}^{R_2}+\frac{1}{2}(n_{i}^{R_1}n_{j}^{R_1}+n_{i}^{R_2}n_{j}^{R_2})\bigg]
\end{aligned}
\label{hamiltonian}
\end{equation}
where $\sigma_{i}^{gr}$ ($r={R_1,R_2}$) represents the $i$-th atom transition between the ground state $\left| g \right\rangle$ and the Rydberg state $\left|  r \right\rangle$, $n_{i}^{R_1,R_2}$ are the population operators for the two Rydberg energy levels $\left|  R_1 \right\rangle$, and $\left|  R_2 \right\rangle$, and $V_{ij}$ are the interactions between the Rydberg atoms located in $\mathbf{r}_i$ and $\mathbf{r}_j$ [through the van der Waals interaction $V_{ij} = C_6 /\left|\mathbf{r}_i-\mathbf{r}_j\right|^6$]. The Lindblad jump terms are given by $\mathcal{L}_r = (\gamma_{r}/2) \sum_i (\hat{\sigma}_i^{r g} \hat{\rho} \hat{\sigma}^{ gr}_i - \{\hat{n}_i^{r},\hat{\rho}\})$, which represents the decay process from the Rydberg state $\left| r \right\rangle$  ($r={R_1,R_2}$) to the ground state $\left| g \right\rangle$. Using the mean-field treatment, we calculate the master equation $\partial_t \hat{\rho} = i [\hat{H},\hat{\rho}] + \mathcal{L}_{R_1}[\hat{\rho}] + \mathcal{L}_{R_2}[\hat{\rho}]$ and obtain the phase diagram of the matrix elements for $\rho_{R_1R_1}(t)$, see more details in Method Section. 

When driven at a single frequency, the system's discrete-time translation symmetry is broken, and its response exhibits $\mathbb{Z}_m$-symmetry, forming a DTC ($m$ = 2) or a high-order DTC ($m$ $>$ 2). Under a dual-frequency quasi-periodic drive, the system's Hamiltonian inherits the same quasi-periodicity, and its response then exhibits a more complex $\mathbb{Z}_m\times\mathbb{Z}_n$-symmetry. If the observable effects in a system are solely those of a $\mathbb{Z}_m$-symmetry, then it is impossible for the system to intrinsically exhibit a $\mathbb{Z}_m\times\mathbb{Z}_n$-symmetry, see examples in Supplementary materials. The subharmonic response of the Rydberg atoms occurs because of the presence of the interaction between Rydberg atoms. Under specific conditions, the system enters a DTQC phase, characterized by quasi-periodic subharmonic responses \cite{zaletel2023colloquium}.

In experiment, we use a three-photon EIT scheme to excite and detect Rydberg atoms, the energy level diagram and the experimental setup are shown in Figs.~\ref{setup}(a) and (b), see more details in Method section. The measured Rydberg excitation non-equilibrium dynamics in the Fourier spectrum manifests a series of peaks at frequencies:
\begin{equation}
\begin{aligned}
f = \frac{N_1}{m}  f_1 + \frac{N_2}{n}  f_2, \quad N_1, N_2 \in \mathbb{Z},  \frac{N_1}{m} ,  \frac{N_2}{n} \notin \mathbb{Z}.
\end{aligned}
\end{equation}
The definitive evidence for a DTQC phase requires that its characteristic subharmonic response remains robust against perturbations of the system's parameters. When distinct values are chosen for \textit{m} and \textit{n}, the resulting subharmonic response reflects different finite Abelian group symmetries. 

Under a two-tone drive with commensurate frequencies $f_1/f_2 \in \mathbb{Q}$ (where $\mathbb{Q}$ denotes the set of rational numbers), the resulting cyclic structure  constrains the system's dynamics to be generated by a single fundamental frequency and its harmonics. When the greatest common divisor of \textit{m} and \textit{n} is unity, i.e.,   greatest common diviso $\gcd(m,n) = 1$ (\textit{m} and \textit{n} are mutually prime), the symmetry group $\mathbb{Z}_m\times\mathbb{Z}_n$ becomes cyclic and is isomorphic to $\mathbb{Z}_{mn}$. In this case, the system's response frequencies lock onto a rational multiple of the drive frequencies. As a result, the dynamics exhibit strict periodicity with a period that is an integer multiple of the driving period, corresponding to a characteristic of a high-order DTC \cite{liu2024higher}. 

In contrast, under incommensurate driving with $f_1/f_2 \notin \mathbb{Q}$ , if \textit{m} and \textit{n} are coprime, the cyclic group symmetry $\mathbb{Z}_{mn}$ still enforces a periodic constraint, preventing the emergence of quasi-periodic order. However, when \textit{m} and \textit{n} are not coprime (for example, when \textit{m} = \textit{n}), this constraint is relaxed, allowing the system to exhibit quasi-periodic dynamics, see the results in Fig.~\ref{setup}(c). This diversity in group symmetry profoundly affects both the robustness and the spectral structure of the subharmonic response, making it a key indicator of different non-equilibrium phases. 

\begin{figure*}
\centering
\includegraphics[width=2.08\columnwidth]{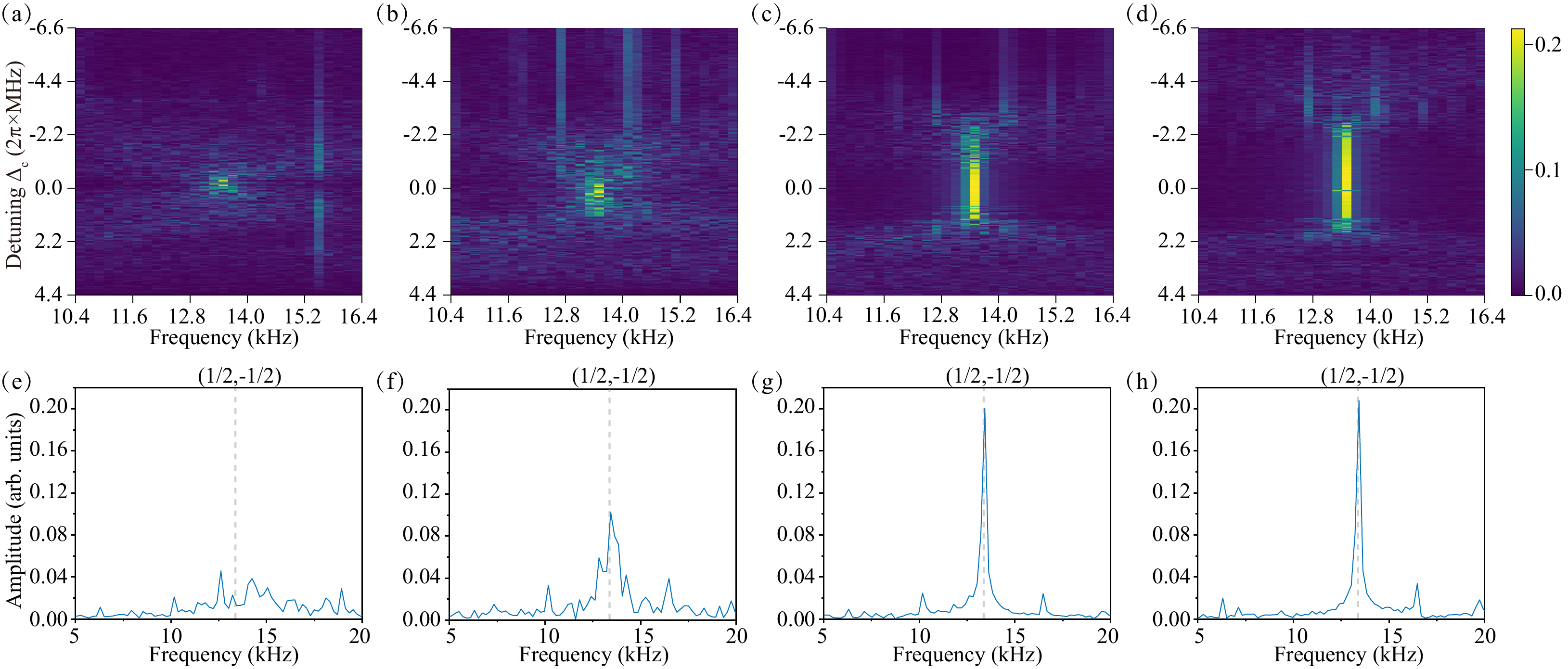}\\
\caption{\textbf{The phase diagram of DTQC versus $\Delta_c$ and the voltage of RF fields.}  When the modulation frequency of $f_1$-field is set to 70 kHz and that of $f_2$-field to 43.262 kHz, we maintain the voltage amplitude of the RF field of $f_2$-field at 1.5 V while gradually increasing the voltage amplitude of $f_1$-field. The frequency of the two RF fields is 7.2 MHz. (a)-(d) show the phase diagrams of the subharmonic response at the position (1/2, -1/2), where the corresponding voltage amplitudes of $f_1$-field are 1.0 V, 1.2 V, 1.4 V, and 1.6 V, respectively. The color bar represents the transmission intensity. (e)-(h) correspond to the Fourier spectrum obtained when the coupling detuning $\Delta _{c}$ = 0 under the above conditions.}
\label{Robustness}
\end{figure*}

\subsection*{Phase diagram}
Under periodic driving by the external RF field, the system is driven out of equilibrium, while the interactions between Rydberg atoms lead to complex subharmonic responses in the system dynamics. Here, we employ two RF fields with dual-frequency periodic modulation. By scanning the coupling detuning $\Delta _{c}$ from $-2\pi \times 17.6$ MHz to $2\pi \times 13.2$ MHz, we plot the phase diagram of the system response by measuring the Fourier spectral of probe transmission. Figure ~\ref{phase diagram}(a) shows the result with only $f_1$-field open, where the applied RF field has a modulation frequency of $f_{1} $ = 70 kHz. It can be observed that within a certain range of detuning $\Delta _{c}$, the system not only responds at the driving frequency but also exhibits oscillations in the probe light transmission at half the driving frequency. This behavior signifies the emergence of the DTC phase, exhibiting $\mathbb{Z}_{2}$-symmetry. Similarly, when only $f_2$-field is activated with a modulation frequency of $f_{2} $ = 43.262 kHz, the system also displays DTC characteristics within a specific detuning range, as shown in Fig.~\ref{phase diagram}(b).

When the two RF fields are simultaneously applied, the results are shown in Fig.~\ref{phase diagram}(c). The ratio between the two frequencies is given by $f_{1} /f_{2}=(\sqrt{5}+1 )/2\approx 1.618$. The incommensurate driving frequencies lead to a quasi-periodic drive. It can be observed that the initial response of the system occurs at the same frequencies as the driving RF fields, indicating that the discrete time-translational symmetry remains unbroken at this stage. As the detuning $\Delta _{c}$ increases, the system gradually exhibits subharmonic responses composed of half-integer frequency combinations of the driving fields, such as $f$ = $f_{1}/2 -f_{2}/2$, $3f_{1}/2 -3f_{2}/2$, and $f_{1}/2 +f_{2}/2$, displaying  $\mathbb{Z}_2\times\mathbb{Z}_2$-symmetry. 

These quasiperiodic subharmonic responses indicate the breakdown of discrete time-translation symmetry under quasi-periodic driving. The system exhibits characteristics of a DTQC, displaying short-range disorder while maintaining long-range order in the time domain. This emergent symmetry can be reconstructed from $\mathbb{Z}_2$-symmetric DTC with $f_{1}$ driving and $\mathbb{Z}_2$-symmetric DTC with $f_{2}$ driving. If one of the driving fields does not excite $\mathbb{Z}_2$-symmetry of system, a temporal quasicrystal cannot be formed. Figures ~\ref{phase diagram}(d)–(f) present the Fourier spectra of the system response under the condition of the detuning $\Delta _{c} =-2\pi \times 8.8$ MHz. It is observed that upon the emergence of the quasicrystal phase, the signature of the DTC phases associated with the individual drives at $f_{1} $ and $f_{2} $ vanishes.
\begin{figure*}
\centering
\includegraphics[width=2.08\columnwidth]{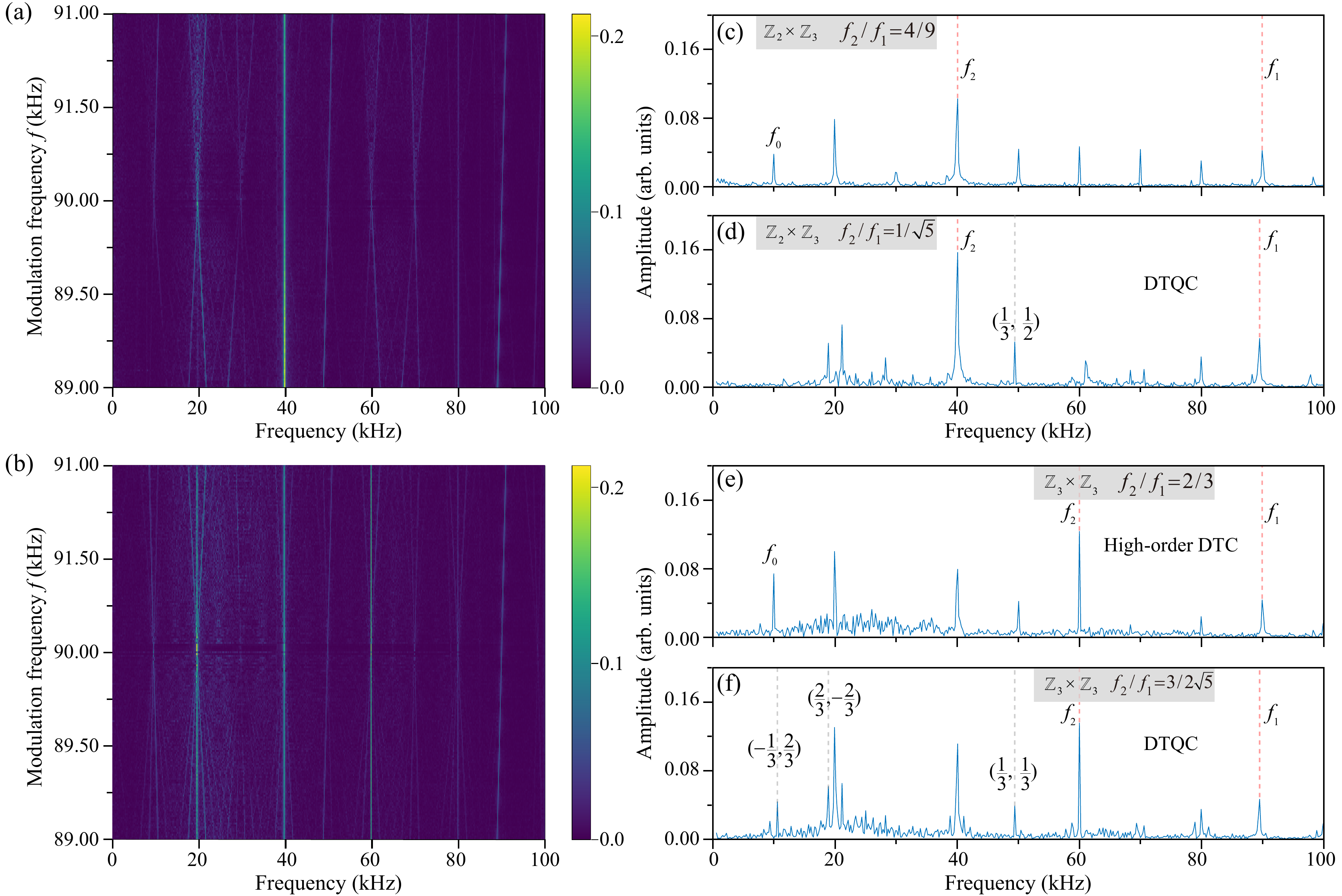}
\caption{\textbf{$\mathbb{Z}_m\times\mathbb{Z}_n$ discrete time quasicrystal (DTQC).} The phase diagrams represent the Fourier spectra of the system response measured under different modulation frequencies. The frequency of RF $f_1$-field is 8.3 MHz with amplitude $U_1$ = 1.7 V, and the Fourier spectra are obtained by sweeping its modulation frequency $f_1$ from 89 kHz to 91 kHz. In panel (a), the frequency of RF $f_2$-field is 8.5 MHz with the amplitude $U_2$ = 1.76 V, and a fixed modulation frequency $f_2$ = 40 kHz. Panel (b) corresponds to the RF $f_2$-field with the frequency of 8.2 MHz, $U_2$ = 1.5 V, and $f_2$= 60 kHz. The color bar represents the transmission intensity. Panel (c) shows the Fourier spectrum when $f_1$ = 90 kHz, $f_2$ = 40 kHz, corresponding to a frequency ratio $f_2$/$f_1$ = 4/9. Panel (d) presents the system response when $f_1$ = $40\sqrt{5}$ $\approx$ 89.443 kHz, illustrating a DTQC with $\mathbb{Z}_2\times\mathbb{Z}_3$-symmetry. Panel (e) displays the Fourier spectrum at $f_1$ = 90 kHz, $f_2$ = 60 kHz with $f_2$/$f_1$ = 2/3. Panel (f) shows the spectrum for $f_1$ = $40\sqrt{5}$ kHz, $f_2$ = 60 kHz and $f_2$/$f_1$ = 3/(2$\sqrt{5}$), including multiple DTQC phases with $\mathbb{Z}_3\times\mathbb{Z}_3$-symmetry marked by grey dotted lines.}
\label{different symmetries}
\end{figure*}

\subsection*{Rigidity of DTQC}
To investigate the effect of the RF field amplitude on the DTQC and to verify the robustness of the DTQC against the perturbations of laser detuning, we varied the voltage of the RF field and measured the corresponding phase diagram of the system, as shown in Fig. \ref{Robustness}. The measured phase diagram reveals the different scenarios of the transmission spectrum as the RF voltage $U$ changes. By adjusting the electrode voltage to modulate the RF field intensity, we measured the Fourier spectrum of the probe laser transmission as a function of the coupling detuning $\Delta _{c}$. Figures \ref{Robustness}(a)-(d) show the phase diagrams of the probe light transmission obtained by scanning $\Delta _{c} =-2\pi \times 6.6$ MHz to $2\pi \times 4.4$ MHz, while the voltage of $f_2$-field is maintained at $U_{2}$ = 1.5 V and varying the voltage of $f_1$-field $U_{1} $ to 1.0 V, 1.2 V, 1.4 V, and 1.6 V, respectively. 

Figures \ref{Robustness}(e)-(h) display examples of the Fourier spectra of the probe field transmission at a coupling detuning of $\Delta _{c}$ = 0 under different voltages $U_{1} $. The phase diagrams demonstrate the robustness of the quasicrystal against perturbations, as evidenced by their stability against small variations in the detuning $\Delta _{c}$. As $\Delta _{c}$ increases or decreases beyond a certain range, the DTQC becomes unstable to minor variations in detuning, leading to its dispersion or splitting into two frequencies symmetric about its original frequency. Moreover, the results indicate that as the intensity of the RF field increases, the strength of the DTQC correspondingly enhances, and the range of its emergence gradually expands, exhibiting that DTQC has become more robust to system parameters.
\begin{figure*}
\centering
\includegraphics[width=2.08\columnwidth]{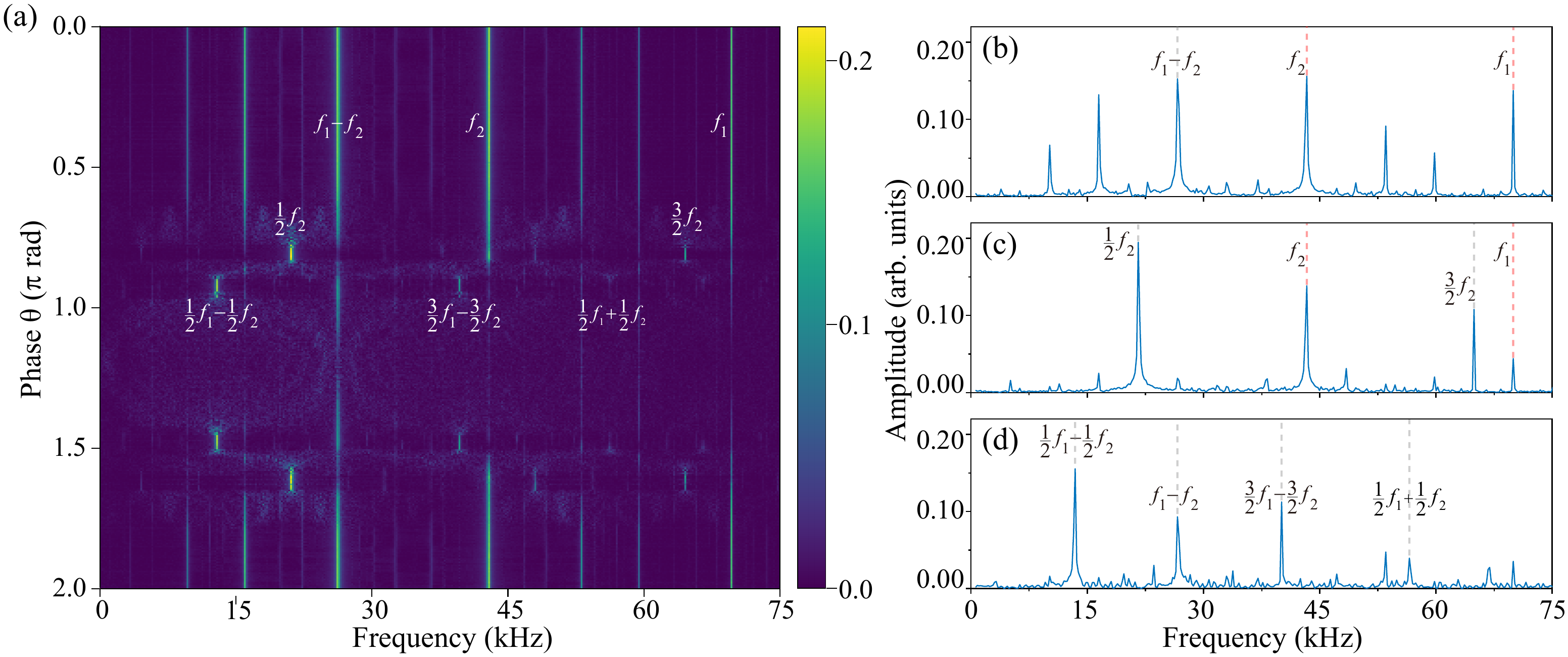}\\
\caption{\textbf{Measured phase diagram versus the relative phase between two RF fields.} The Fourier spectral phase diagram is obtained by measuring the system response while changing the phase of the RF fields. The voltage of $f_1$-field is set to $U_{1} $ = 1.7 V, modulation frequency $f_{1} $ = 70 kHz. The voltage of $f_2$-field is $U_{2} $ = 1.5 V, modulation frequency $f_{2} $ = 43.262 kHz; the frequency of both RF fields is 7.2 MHz. (a) Measured Fourier spectra as a function of the relative phase $\theta $ between the $f_1$-field and $f_2$-field. The carrier phase of $f_2$-field is fixed at 0, while that of $f_1$-field is scanned from 0 to 2$\pi $. The color bar represents transmission intensity. (b)-(d) show the corresponding Fourier spectra at relative phases of $\theta $ = 90°(b), $\theta $ = 145°(c), and $\theta $ =165°(d), respectively. In (c), the first peak frequency $f=f_{2} /2$ corresponds to the second-order DTC of $f_{2} $; in (d), the frequency $f=f_{1} /2-f_{2} /2$ corresponds to the DTQC.}
\label{phase transition}
\end{figure*}

\subsection*{$\mathbb{Z}_m\times\mathbb{Z}_n$ DTQC}
As discussed above, the interplay between the commensurability of the two driving frequencies and the coprimality of \textit{m} and \textit{n} fundamentally determines the emergent symmetries and the resulting non-equilibrium dynamics of the system. Studying $\mathbb{Z}_m\times\mathbb{Z}_n$ temporal quasicrystals is therefore essential to uncover how symmetry governs the transition between periodic and quasi-periodic phases in driven systems. We consider different experimental conditions to investigate these.

Figures  \ref{different symmetries}(a) and (b) show the measured Fourier phase diagrams, the rich DTQC phases can be clearly observed under different symmetry and frequency-commensurability conditions. In Fig.  \ref{different symmetries}(c), with $\mathbb{Z}_2\times\mathbb{Z}_3$-symmetry and a commensurate frequency ratio $f_2$/$f_1$ = 4/9, the measured Fourier spectra seem normal and display several regular peaks. The spectrum structure is dominated by a fundamental frequency and its harmonics, reflecting strict periodicity as $\gcd(f_1,f_2) = 10$ kHz. When we consider a more complex case $f_2$/$f_1$ = 5/9 ($\gcd(f_1,f_2) = 10$ kHz), the signature of DTC phase at frequency $f_0$ = 5 kHz can be observed and periodic temporal order locking effect appears due to the cyclic nature of the effective symmetry group $\mathbb{Z}_6$ (isomorphic to $\mathbb{Z}_2\times\mathbb{Z}_3$), see the measured results in Supplementary materials.

In contrast, when we consider a case of incommensurate frequencies driving, Fig. \ref{different symmetries}(d) shows the same $\mathbb{Z}_2\times\mathbb{Z}_3$-symmetry but with an incommensurate drive ratio $f_2$/$f_1$ = 1/$\sqrt{5}$. In this case, the Fourier spectrum displays incommensurate frequency response, indicating a breakdown of periodic locking and the formation of a DTQC at frequency $f_{1}/3 +f_{2}/2$. While, due to the presence of the cyclic group symmetry $\mathbb{Z}_{6}$, it still enforces a periodic constraint [see the suppressed response at the frequency of $f_{2}/2$], preventing the emergence of quasi-periodic order in the range of low frequency. This result demonstrates how the quasiperiodic temporal order survives under the strong constraint of the $\mathbb{Z}_2\times\mathbb{Z}_3$-symmetry.

Similarly, in Fig. \ref{different symmetries}(e), under $\mathbb{Z}_3\times\mathbb{Z}_3$-symmetry and a rational frequency ratio $f_2$/$f_1$ = 2/3 and $\gcd(f_1,f_2) = 30$ kHz, the system forms a high-order DTC because the symmetry of the system is formed by two same $\mathbb{Z}_3$-symmetry. The system response displays a minimal single frequency at $f_0 = $ $f_{1}/3 - f_{2}/3$ = 10 kHz due to the interplay between $\mathbb{Z}_3\times\mathbb{Z}_3$-symmetry and commensurability. The Fourier spectrum shows clear sharp harmonic peaks consistent with periodic dynamics. Meanwhile, in Fig. \ref{different symmetries}(f), with the same symmetry but an incommensurate ratio $f_2$/$f_1$ = 3/2$\sqrt{5}$, the dynamics become quasiperiodic. In this case, the periodic constraint is relaxed (see the subharmonic response at the frequency of $f_{2}/3$) and the system response at $f_2$/3 can be observed. The spectrum contains numerous non-commensurate peaks, for example $-f_{1}/3 +2f_{2}/3$, $2f_{1}/3 -2f_{2}/3$, $f_{1}/3 +f_{2}/3$, confirming the presence of several DTQC phases with rich spectral structure and without long-term periodic recurrence.

\subsection*{Phase-dependent DTQC}
To investigate the more feature of the quasicrystalline phase in the system, we further adjust the relevant parameters of the RF fields and measured the Fourier spectrum of the probe field transmission. By varying the relative phase $\theta $ of the applied RF field, we maintain the phase of one RF field constant while altering the phase of the other field, thereby inducing changes in the relative phase between the two RF fields. During the experiment, while maintaining a fixed laser coupling detuning $\Delta _{c}$, we set the phase of $f_2$-field fixed at $\theta _{2} $ = 0, we scan the phase $\theta _{1} $ of $f_1$-field from 0 to $2\pi $ and observe the phase diagram of the system response, as shown in Fig. \ref{phase transition}. In Fig. \ref{phase transition}(a), we observe phase transitions from the DTC to the DTQC and from the DTQC to the DTC in the system response. As the relative phase $\theta $ gradually increases from 0, the system exhibits harmonic responses. Figure \ref{phase transition}(b) displays the Fourier spectrum of the probe light at $\theta $ = 90° as an example. 

With further increase in the relative phase between the RF fields, the system response becomes subharmonic relative to the driving frequency, manifesting as the frequency responses at $f_{2} /2$ and $3f_{2} /2$, respectively, indicating the system enter into the DTC phase driven at frequency $f_{2}$. Figure \ref{phase transition}(c) plots the Fourier spectrum at $\theta $ = 145°, where the first labeled peak corresponds to the $n$-DTC phase with $n$ = 2. As we further increase the value of $\theta $, the system undergoes a phase transition, switching from the DTC phase to the DTQC phase. The Fourier spectrum at this stage, exemplified by $\theta $ = 165° in Fig. \ref{phase transition}(d), reveals combined frequency responses at half-frequencies between the driving frequencies $f_{1}$ and $f_{2}$, such as $f_{1}/2-f_{2}/2$, $3f_{1}/2-3f_{2}/2$, and $f_{1}/2 +f_{2}/2$. The rich behavior and phase transitions observed as a function of the relative RF phase $\theta $ arise from the competition between two non-commensurate collective excitation processes, each attempting to impose its own temporal order on the system.

\section*{DISCUSSION}
We have experimentally observed different robust DTQCs and rich phases in a quasiperiodically driven Rydberg atomic gas. The interaction between the driving field and the Rydberg atoms leads to the emergence of rich discrete time-translation symmetry breaking. A phase transition from DTC to DTQC was observed, in which the radio-frequency field plays a crucial role in regulating the behavior of complex many-body systems. The observed spectral signature of the DTQC in $\mathbb{Z}_m\times\mathbb{Z}_n$-symmetry reflects a fractal-like frequency structure that is more universal characteristic of quasiperiodic systems. 

In the experiment, the long-range van der Waals interactions between Rydberg atoms are central to the emergence of DTQCs in this system.  Without this long-range interaction, the system would behave as a collection of independent atoms, each responding linearly to the drive, resulting in oscillations only at the fundamental driving frequencies and their immediate harmonics. The presence of interactions, however, generates higher-order correlations and non-linear feedback, which mix the two incommensurate driving frequencies. This mixing produces robust, incommensurate subharmonic peaks in the Fourier spectrum, such as DTQCs at frequencies of $f_1/2$-$f_2/2$ or $f_1/2$+$f_2/3$ .

The study of DTQCs in Rydberg atomic systems advances the theory of DTQCs and opens new avenues for exploring diverse manifestations of time-translation symmetry breaking in driven quantum systems. Future studies may investigate the symmetry-constrained non-equilibrium dynamics, as well as the role of long-range interactions in stabilizing such phases. This work provides a platform to engineer more complex finite Abelian symmetries beyond $\mathbb{Z}_2\times\mathbb{Z}_2$ opens the door to realizing custom temporal patterns (such as quasicrystals with higher-order symmetries) in a highly controllable quantum system. 

\begin{figure*}
\centering
\includegraphics[width=2.08\columnwidth]{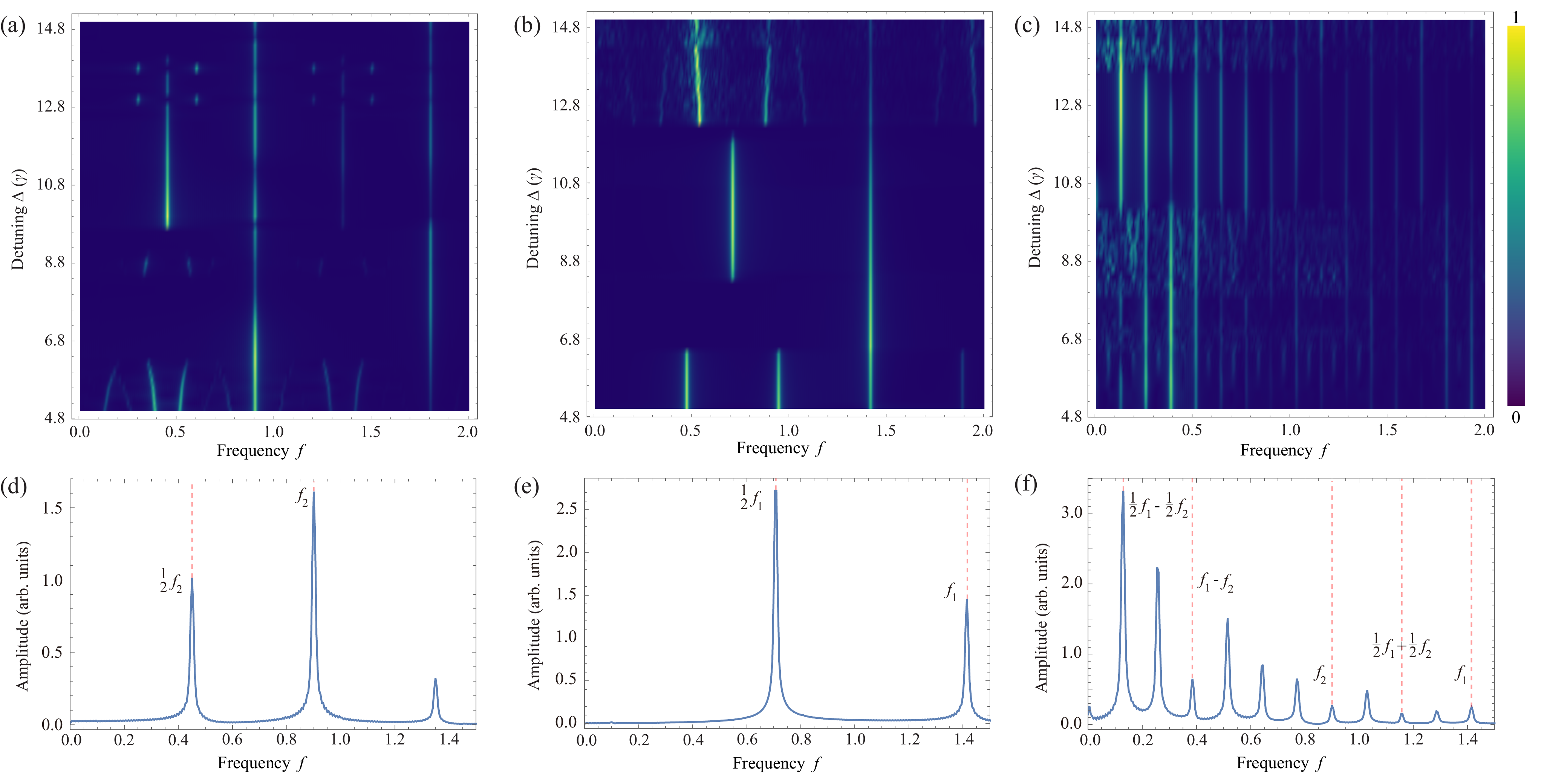}\\
\caption{\textbf{Theoretically simulated phase diagrams.} Calculated Fourier spectra of the Rydberg atom population $\rho_{R_1R_1}$ with parameters $V$ = -16$\gamma$, $\Omega$ = 3.5$\gamma$, $\delta$ = 12$\gamma$. (a), (b) and (c) correspond to the results of single-frequency drive with $f_2 = 0.9\gamma$, single-frequency drive with $f_1 = \sqrt{2}\gamma$, and dual-frequency drive with both $f_1$ and $f_2$ simultaneously present, respectively. (d) ,(e) and (f) correspond to the Fourier spectra of the system response under different driving conditions, where the parameters are $\Delta_{f_1}$ = -10.8$\gamma$ and $\Delta_{f_2}$ = -10$\gamma$. The color bar represents the Fourier transform intensity.}
\label{Numerical results}
\end{figure*}

\section*{Methods}

\subsection*{Experimental setup}
In the experiments, we utilize thermal Cesium atoms to investigate the properties of DTQC. A three-photon EIT scheme is used to prepare and detect Rydberg atoms. The Cesium energy level structure and the experimental setup  are shown in Figs. \ref{setup}(a) and (b), respectively.  Specifically, the excitation process involves: using an 852 $nm$ probe beam with a Rabi frequency $\Omega _{p}$ to resonantly driving the transition from state $\left | 6S_{1/2}  \right \rangle $ to state $\left | 6P_{3/2}   \right \rangle $; a resonant 1470 $nm$ laser with a Rabi frequency $\Omega _{d}$ to resonantly driving the transition from state $\left | 6P_{3/2}   \right \rangle $ to state $\left | 7S_{1/2 }   \right \rangle $, and a 780 $nm$ coupling beam with a Rabi frequency $\Omega _{c}$ and a detuning $\Delta _{c}$  to driving the transition from state $\left | 7S_{1/2 }   \right \rangle $ to state $\left | nP   \right \rangle $. The atoms are irradiated with RF fields. 

An 852 $nm$ external cavity diode laser (ECDL) is frequency-stabilized using saturation absorption spectroscopy, while a second ECDL at 1470 nm is locked via two-photon spectroscopy. The 780 nm laser, amplified by a tapered amplifier, serves as the coupling beam. The probe laser counter-propagates relative to the dressing and coupling beams. Two RF signals are generated by an arbitrary function generator (AFG, Rigol DG4000 series); their outputs are combined via coaxial cables and delivered to a pair of electrode plates. The transmitted probe light is detected by a photodetector and recorded on an oscilloscope. To scan external parameters and map out the phase diagram of the system response, the coupling laser, oscilloscope, and AFG are all synchronized under computer control.

The probe, dressing, and coupling beams are aligned to counter-propagate at a small angle and pass in parallel through a 7 cm-long vapor cell. The probe light is focused on the vapor cell with a beam waist radius $(1/e^{2} )$ of approximately 200 $\mu m$ and an intensity of 64 $\mu W$. The dressing and coupling lights are focused to beam waists of about 500 $\mu m$, with powers of 16.8 mW and 1.5 W, respectively. The corresponding Rabi frequencies are $\Omega _{p}=2\pi\times 35$ MHz, $\Omega _{d}=2\pi\times 14$ MHz, and $\Omega _{c}=2\pi\times42$ MHz, respectively. A pair of circular copper electrode plates, each 3 mm thick and 120 mm in diameter, are mounted parallel to each other with a separation of 40 mm. The Cesium vapor cell is positioned at the center between these electrodes.

\subsection*{Numerical results}
Due to the thermal motion of atoms, we can neglect the correlations between atoms and thus employ the mean-field approximation\cite{liu2024higher,ding2023ergodicity,carr2013nonequilibrium,wu2023observation}. For the three-level system considered in the theoretical model, which includes a ground state and two Rydberg states, based on the system's Hamiltonian Eq.\ref{hamiltonian}, we obtain the master equation of the system under the mean-field approximation as follows:
\begin{equation}
\begin{aligned}
\frac{\partial}{\partial t} \rho_{R_1 R_1} & =i \frac{\Omega}{2}\left(\rho_{g R_1}-\rho_{R_1 g}\right)-\gamma \rho_{R_1 R_1}, \\
\frac{\partial}{\partial t} \rho_{R_2 R_2} & =i \frac{\Omega}{2}\left(\rho_{g R_2}-\rho_{R_2 g}\right)-\gamma \rho_{R_2 R_2}, \\
\frac{\partial}{\partial t} \rho_{g R_1} & =i \frac{\Omega}{2}\left(\rho_{R_1 R_1}+\rho_{R_2 R_1}-\rho_{g g}\right) \\ 
& +i\left((\Delta_{f_1}(t)+\Delta_{f_2}(t) )-V_{\rm{MF}}+i \frac{\gamma}{2}\right) \rho_{g R_1}, \\
\frac{\partial}{\partial t} \rho_{g R_2} & =i \frac{\Omega}{2}\left(\rho_{R_2 R_2}+\rho_{R_1 R_2}-\rho_{g g}\right)\\
& +i\left((\Delta_{f_1}(t)+\Delta_{f_2}(t) )+\delta-V_{\rm{MF}}+i \frac{\gamma}{2}\right) \rho_{g R_2}, \\
\frac{\partial}{\partial t} \rho_{R_1 R_2} & =i \frac{\Omega}{2}\left(\rho_{g R_2}-\rho_{R_1 g}\right)-i\left(\delta-i \gamma\right) \rho_{R_1 R_2},
\end{aligned}
\end{equation}
where $V_{\rm{MF}} = V(\rho_{R_1 R_1}+\rho_{R_2 R_2})$ is the mean field shift, and we set the effective Rabi frequency $\Omega_1=\Omega_2=\Omega$. Here, we treat a simplified model by using same interaction $V_{ij}=V$ by ignoring the difference between different sublevels of Rydberg atoms. $\Delta_{f_1}(t) $ and $\Delta_{f_2}(t) $ correspond to the energy shifts induced by dual-frequency driving respectively. By solving the equations above, we can obtain the time response of the system and can also obtain the Fourier spectrum via discrete Fourier transformation as shown in Fig. \ref{Numerical results}. 

We calculated the system response phase diagrams for single-frequency driving with $f_1 = \sqrt{2}\gamma$, single-frequency driving with $f_2 = 0.9\gamma$, and dual-frequency driving with both $f_1$ and $f_2$ applied simultaneously, as shown in Figs. \ref{Numerical results}(a-c). For single-frequency driving, scanning the system parameters we can observe the presence of 2-DTC and 3-DTC. By applying a dual-frequency quasiperiodic driving when the frequencies $f_1$ and $f_2$ are non-commutative, the system response manifests a series of subharmonic peaks at frequencies, corresponding to discrete-time quasicrystals. Furthermore, we have shown the Fourier spectra of the system response under different driving conditions with the parameters $\Delta_{f_1}$ = -10.8$\gamma$ and $\Delta_{f_2}$ = -10$\gamma$, the results are as illustrated in Figs. \ref{Numerical results}(d-f).

\section*{Acknowledgements}
We thank for the previous discussions with Prof.  Krzysztof Sacha on discrete time quasicrystals. We acknowledge funding from the National Key R and D Program of China (Grant No. 2022YFA1404002), the National Natural Science Foundation of China (Grant Nos. T2495253, 62435018).

\section*{Data Availability}
All experimental data used in this study are available from the corresponding author upon request.

\section*{Author contributions statement}
D.-S.D. and B.L. conceived the idea. D.Y.Z., Z.Y.Z. and Q.F.W. conducted the physical experiments. D.-S.D., B.L. and D.Y.Z wrote the manuscript. The research was supervised by D.-S.D. All authors contributed to discussions regarding the results and the analysis contained in the manuscript.

\section*{Competing interests}
The authors declare no competing interests.

\bibliography{ref}

\end{document}


\title{Supplementary materials for the “Observation of  Discrete Time Quasicrystal in Rydberg Atomic Gases”}

\author{Dong-Yang Zhu$^{1,2,\textcolor{blue}{\star}}$}
\author{Zheng-Yuan Zhang$^{1,2,\textcolor{blue}{\star}}$}
\author{Qi-Feng Wang$^{1,2,\textcolor{blue}{\star}}$}
\author{Yu Ma$^{1,2}$}
\author{Tian-Yu Han$^{1,2}$}
\author{Chao Yu$^{1,2}$}
\author{Qiao-Qiao Fang$^{1,2}$}
\author{Shi-Yao Shao$^{1,2}$}
\author{Qing Li$^{1,2}$}
\author{Ya-Jun Wang$^{1,2}$}
\author{Jun Zhang$^{1,2}$}
\author{Han-Chao Chen$^{1,2}$}
\author{Xin Liu$^{1,2}$}
\author{Jia-Dou Nan$^{1,2}$}
\author{Yi-Ming Yin$^{1,2}$}
\author{Li-Hua Zhang$^{1,2}$}
\author{Guang-Can Guo$^{1,2}$}
\author{Bang Liu$^{1,2,\textcolor{blue}{\ddagger}}$}
\author{Dong-Sheng Ding$^{1,2,\textcolor{blue}{\dagger}}$}
\author{Bao-Sen Shi$^{1,2}$}

\affiliation{$^1$Laboratory of Quantum Information, University of Science and Technology of China, Hefei, Anhui 230026, China.}
\affiliation{$^2$Anhui Province Key Laboratory of Quantum Network, University of Science and Technology of China, Hefei 230026, China.}

\date{\today}
\symbolfootnote[1]{D.Y.Z, Z.Y.Z, Q.F. W contribute equally to this work.}
\symbolfootnote[3]{lb2016wu@ustc.edu.cn}
\symbolfootnote[2]{dds@ustc.edu.cn}

\maketitle
This supplementary material reports a discrete time crystal (DTC) under commensurate dual-frequency driving and the limitations for emergent symmetries.  

\subsection*{DTC with Commensurate Frequency}
\begin{figure*}
\centering
\includegraphics[width=1.65\columnwidth]{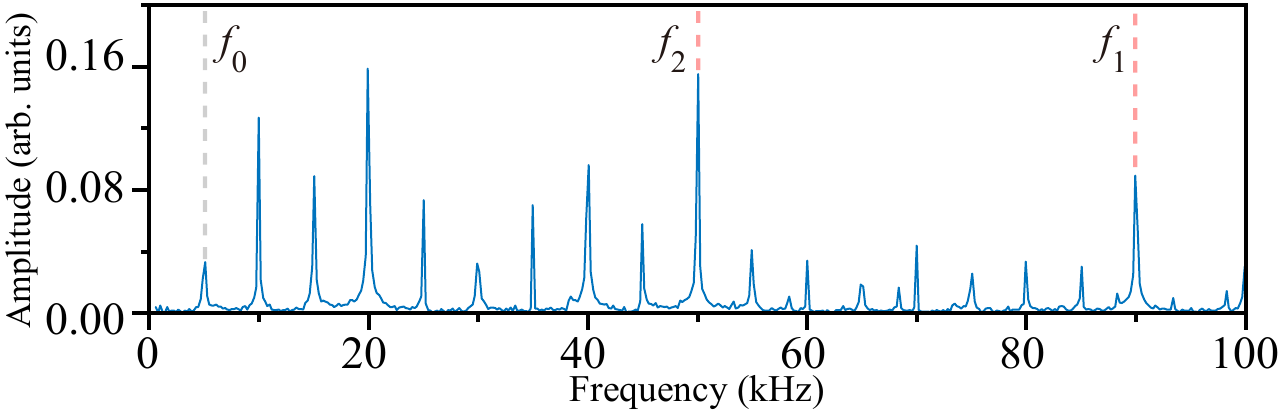}\\
\caption{\textbf{The measured Fourier spectrum. }Here, the carrier frequency of RF $f_1$-field is set to 8.72 MHz, with voltage $U_{1} $ = 1.7 V and modulation frequency $f_{1} $ = 90 kHz. The carrier frequency of RF $f_2$-field is set to 8.00 MHz, with voltage $U_{2} $ = 1.3 V and modulation frequency $f_{2} $ = 50 kHz. The positions of $f_{1} $ and $f_{2} $ are marked with red dashed lines.
}
\label{Appendix1}
\end{figure*}

The measured Fourier spectrum in Fig.~\ref{Appendix1} reveals a clear signature of a DTC phase under specific symmetry and frequency conditions. With RF $f_1$-field modulated at $f_1$ = 90 kHz and RF $f_2$-field at $f_2$ = 50 kHz, the frequency ratio $f_2/f_1$ = 5/9 is commensurate, yielding a greatest common divisor (gcd) frequency of 10 kHz. In this case, the system response oscillates at 10 kHz or its harmonics. However, a fundamental frequency of $f_0$ = 5 kHz emerges, accompanied by a series of subharmonic peaks. This indicates the formation of a DTC phase, where the temporal order is locked at half the gcd frequency of $f_1$ and $f_2$. 

The emergence of the 5 kHz subharmonic can be understood through the effective symmetry group $\mathbb{Z}_6$, which is isomorphic to $\mathbb{Z}_2\times\mathbb{Z}_3$. This group structure arises from the combination of the individual symmetries associated with the two driving fields. The $\mathbb{Z}_2$-symmetry corresponds to the period-doubling behavior induced by the 50 kHz modulation, while the $\mathbb{Z}_3$-symmetry stems from the period-tripling response due to the 90 kHz modulation. Their interplay leads to a composite temporal order characterized by an oscillation at $f_0$ = $f_1$/3 -$f_2$/2 = 5 kHz, reflecting a coherent locking into a lower-frequency collective mode. 

This behavior exemplifies a robust DTC phase, where the system spontaneously breaks the discrete time-translation symmetry of the drive. The observed Fourier spectrum not only confirms the prediction of such phases under commensurate dual-frequency driving (see Eq. (2) in main text), but also highlights the role of synthetic cyclic symmetries in stabilizing non-trivial temporal orders.  

\subsection*{Limitations on Emergent Symmetries}
\begin{figure*}
\centering
\includegraphics[width=1.65\columnwidth]{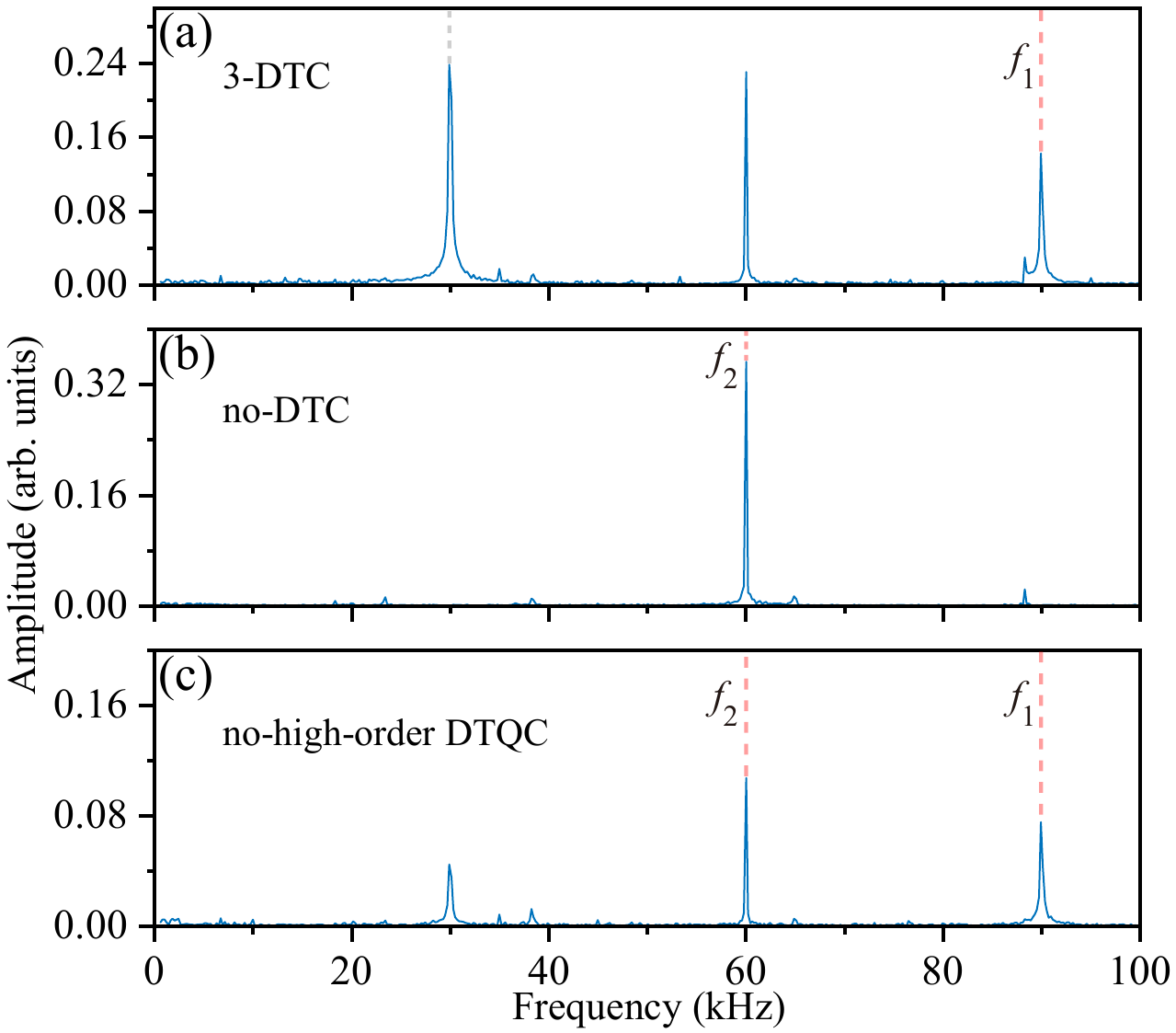}\\
\caption{\textbf{The measured Fourier spectrum.}  Here, the carrier frequency of RF $f_1$-field is set to 8.72 MHz, with a voltage $U_{1} $ = 1.7 V and a modulation frequency $f_{1} $ = 90 kHz. The carrier frequency of RF $f_2$-field is set to 8.20 MHz, with a voltage $U_{1} $ = 1.42 V and a modulation frequency $f_{2} $ = 60 kHz. The positions of $f_{1} $ and $f_{2} $ are indicated by red dashed lines.}
\label{Appendix2}
\end{figure*}

The experimental results presented in the Fourier spectrum illustrate the distinct dynamical responses of the system under different driving conditions, confirming the intrinsic limitations regarding emergent symmetries. When only RF $f_1$-field is applied with a modulation frequency of $f_1$ = 90 kHz, the system exhibits a clear period-tripling response, as evidenced by the prominent peak at $f_0$ = $f_1$/3 in the $m$-DTC case, where $m$ = 3 (see Fig.~\ref{Appendix2}(a)). This signifies the emergence of a discrete time translation symmetry breaking, characteristic of a $\mathbb{Z}_3$ structure, or more generally, a $\mathbb{Z}_3$-symmetric system response, which depends on the specific period multiplication. 

Upon introducing RF $f_2$-field with a modulation frequency of $f_2$ = 60 kHz, while carefully tuning its carrier frequency and voltage to avoid any intrinsic period-tripling (no-DTC in Fig.~\ref{Appendix2}(b)), the system shows no subharmonic response at $f_2$. This indicates the absence of spontaneous symmetry breaking under this driving condition alone. However, when both fields are applied simultaneously (there is no-DTC signal in Fig.~\ref{Appendix2}(c)), no additional peaks corresponding to subharmonics periodic order for two driving frequency or combined symmetry breaking (such as those expected from a $\mathbb{Z}_3\times\mathbb{Z}_3$-symmetry in main text) are observed. The spectrum lacks signatures of commensurate frequencies or their linear combinations, implying that no higher-dimensional symmetry emerges.

This supports the assertion that if the observable dynamics of a system are constrained to those of a $\mathbb{Z}_m$-symmetry, it is impossible for the system to intrinsically realize a $\mathbb{Z}_m\times\mathbb{Z}_n$-symmetry. The absence of a DTC under dual-frequency driving underscores the fact that the emergence of symmetries is not merely additive; instead, it depends fundamentally on the interplay between driving parameters and the intrinsic dynamics of the system. Here, the conditions that permit a single $\mathbb{Z}_m$-symmetric DTC do not facilitate the formation of a larger symmetry group, even under additional drives, highlighting the subtle role of dimensionality and coupling in synthetic time crystals.

